\documentstyle[preprint,tighten,prd,aps]{revtex}

\begin{document} \draft


\title{ Probing String Theory with Modulated Cosmological Fluctuations}
\author{ Lev Kofman}
\address{
  CITA, University of Toronto\\
  Toronto, Ontario, Canada, M5S 3H8\\
  {\rm E-mail: \texttt{kofman@cita.utoronto.ca}}
}
\date{March 2003}
\maketitle

\begin{abstract}
Superstring theory, models with extra dimensions
and other  SUSY models generically predict 
that the coupling constants are in fact vacuum expectation values of
fields like the dilaton, moduli etc. Assuming some of these fields
are light during inflation, we get generation of small classical 
inhomogeneities in these fields from inflation. Consequently,
coupling constants inherit small inhomogeneities 
at scales much larger than the causal horizon in the early universe.
After the moduli get pinned down to their minima, the spatial variations of coupling constants
in the late time universe will be erased. However, 
 inhomogeneities in coupling constants in the very early universe
 would generate modulated large scale 
fluctuations in all relic species that are produced  due to interactions
and freezing out. Moreover (p)reheating of the inflaton field results
in modulated curvature fluctuations. Even if the standard inflaton fluctuations are 
suppressed, in  this picture
we may have pure curvature cosmological fluctuations  entirely generated by
 the modulated spatial variations  of the coupling constants during preheating.

\end{abstract}

\pacs{PACS numbers: 04.50.+h, 98.80.Cq \hfill CITA-2003-12}
\narrowtext


\section{Introduction} \label{sec:intro}

Early universe cosmology extends up to
energy scales where we expect superstring/M theory  to operate.
It is of great interest to understand, even in principle, the potential
signatures of superstring theory  on cosmological observables.

The primordial inflationary stage of the now standard cosmological 
paradigm
acts like a universal  amplifier and stretcher of fluctuations.
Indeed, vacuum fluctuations of all light (minimally coupled)
degrees of freedom are unstable during inflation and appear  in subsequent
cosmological evolution as classical  fluctuations with wavelengths
 on cosmological scales.

In inflationary models in $3+1$ dimensional theory,
a scalar, or curvature
 mode of the metric fluctuations $\Phi$ is generated from 
 quantum fluctuations of the inflaton
 field $\delta \phi$ \cite{scalar}
while quantum production of 
  gravitational waves (gravitons)
results in a tensor mode  of  metric fluctuations $h_{\mu\nu}$ \cite{tensor}.
In the simplest inflationary models both modes have almost scale free power spectra
of fluctuations.
Gravitational waves generated from inflation with
a Hubble parameter $H$ have amplitude $k^{3/2}h_k \simeq H/M_P$,
while the amplitude of the scalar mode depends on the inflaton potential
$V(\phi)$, $k^{3/2}\Phi_k \simeq \frac{V^{3/2}}{M_P^2 V_{,\phi}}$.  In 
simple inflationary models, including chaotic inflation or hybrid inflation, the amplitude 
of the scalar mode is 
larger than that of the tensor mode by a factor of ten or more.

The dominant scalar mode  and subdominant tensor mode  both contribute to
 the observed CMB temperature anisotropy with magnitude 
$\frac{\Delta T}{T}
\sim 10^{-5}$,
  while only the scalar mode contributes to large scale matter clustering.
Independently, the tensor mode generate a very small 
transversal-traceless (B) component of CMB polarization anisotropy. The 
ever-growing precision of CMB anisotropy measurments in combination with
large scale structure observations will open up the possibility of testing 
effects
in cosmological fluctuations at the next,  more precise level.
 I will try to motivate consideration of isocurvature
cosmological fluctuations at this level. As an illustration we may use a
parallel with basic atomic spectra and their fine structure.

 In the context of inflationary cosmological fluctuations, one area of
recent interest in string cosmology has been trans-planckian effects which
could cause very small changes in primordial cosmological fluctuations at
very small level if any at all.
 Cosmologists working with the braneworlds were looking
for the features of extra dimensions in the $T/S$ ratio and
details of the power spectra.  Notice,
however, that similar effects with respect to individual tests (say, 
the ratio $T/S$) may be attributed to other choices of inflationary models or 
other
(tuned) effects during inflation.

In this paper I suggest a rather different  idea \cite{cosmo} of how string/brane cosmology
physics may enter  in the theory of   cosmological fluctuations. 
Some of the  effects we describe here
may be tested observationally at a magnitude
 of the order of the gravitational wave amplitude
$\sim  {H \over M_P}$ or even higher, depending on a
model-dependent numerical pre-factor.         
The main suggestion, however, will be 
 a new mechanism of generatation of almost scale free
curvature fluctuations. which we will call modulated curvature fluctuations
for reasons explained below. This opens a new possibilities to
construct the amplitude and the spectra of primordial scalar fluctuations,
rather detached from the inflation model building.

Superstring/M theory, supergravity and many 
phenomenological models of  high energy physics are formulated in
various  numbers of dimensions $D > 4$.  The extra dimensions of  fundamental theories 
have their most obvious relevance in application to cosmology. 
 In effective four dimensional theory, 
 compactified extra dimensions emerge as
scalar fields,  moduli, which we denote
collectively as $\chi_a$.
Scalar fields also arise from bulk scalar fields like the dilaton
and from projections of other fields. 
 These moduli fields must be stabilized by some mechanism, although 
moduli with runaway potentials are not excluded.
The fields $\chi_a$ are thought to   acquire masses $m_a$ 
due to SUSY breaking  sometime in the early universe.

Moduli fields in cosmology pose a problem.
 If the high energy and low energy moduli potentials are different,
then after inflation the moduli will acquire large amplitudes.
 Residual moduli field 
oscillations will have a matter equation of state with energy density $\varepsilon 
\propto 1/a^3$, and they will thus quickly become dominant, before Big Bang Nucleosynthesis.
 Even if this problem is resolved,
one encounters another problem. Moduli couple to matter gravitationally, 
so their time  decay  $\tau \sim \Gamma^{-1} \simeq 
\frac{M_P^2}{m_{\chi}^3} \sim 10^5 (\frac{{\text Tev}}{m_{\chi}})^3 $ seconds.
The products of ${\text Tev }$ mass scale moduli decay may destroy successful BBN.

 We will dare to ignore the dangerous effects of moduli on cosmology,
 assuming they are somehow resolved\footnote{This could be done, for instance, by choosing large 
$m_a$, or by choosing a larger gravitational coupling as in theories
with large extra dimensions, or by allowing runaway moduli 
without stabilization, see \cite{dine}. }, and focus instead on potentially useful
 consequences of these fields.
The  mass scale  $m_a$ and the SUSY breaking energy scale  are
 smaller than the
energy scale of inflation $M \sim 10^{16}$ Gev, $V \simeq M^4$, $H^2 \sim V/M_P^2$,
 $H \sim 10^{13}$ Gev.
As long as the moduli fields or dilaton are lighter that  $10^{13}$ Gev, the  outer space
inflation can generate long-wavelength classical spatial
inhomogeneities of the moduli
fields $\delta \chi_i \sim H$. We do not expect any impact of moduli
inhomogeneities after they are pinned down to their minima at some time $t_p$.

However,  spatial inhomogeneities in the moduli fields present 
 between the end of inflation and BBN may have significant impact if they
 are imprinted in the coupling constants $\alpha_i$, making them spatially varying
at  scales larger than  the casual horizon, $\frac{\delta \alpha}{\alpha} \sim
\frac{\delta \chi}{\chi}\sim \frac{H}{M_P}$. 
Such inhomogeneities would generate large scale isocurvature cosmological fluctuations
in all components that are generated due to interactions  and freezing out,
including radiation, cold dark matter, baryons, etc. before $t_p$.
We will call these modulated  fluctuations.

A more radical extension of this idea is to have not only modulated isocurvature
but modulated curvature fluctuations. Indeed, if scalar fluctuations
of the inflaton field
are suppressed, moduli fluctuations may generate significant fluctuations
in the matter created during (p)reheating after inflation by
spatial modulation of the coupling constants between inflaton and matter.

\section{Moduli fluctuations from inflation}\label{sec:intro}

To establish this story, consider the pure gravity sector of low energy string theory
$S=\frac{1}{16\pi \kappa_D^2} \int d^D x \, e^{-2\Phi}\sqrt{-G}\, {\cal R}$.
The usual (toy model) dimensional reduction from $D$ dimensional space $(x^{\mu}, y^i)$
to  $3+1$ space-time $x^{\mu}$ 
gives us (e.g., \cite{strings})
\begin{equation}\label{action}
S=\frac{1}{16\pi \kappa_4^2} \int d^4 x \, \sqrt{-g}\, \left(R +\frac{1}{2}\chi_a\chi^a
+ {\cal L}_m \right) \ .
\end{equation}
Here we collectively denoted moduli fields as $\chi_a$.
For instance, the graviton in the $D$ dimensional bulk has 
${1 \over 2}D(D-3)$ degrees of freedom.  The
massless $3+1$ dimensional KK
projection contain two degrees of freedom in the usual
tensor mode, $2(D-4)$ gravi-vector degrees of freedom 
and ${1 \over 2} (D-3)(D-4)$ gravi-scalars.
The matter lagrangian ${\cal L}_m$ also  contains  moduli fields
(including the dilaton).
Most important for us will be the fact that the coupling constant in $\Lambda_m$
depends on $\chi_a$ \cite{strings}. 
We will have similar construction in the braneworld scenario, where 
one of the $\chi_a$ will be a radion, coupling with  matter at the brane \cite{brane}.

Suppose  $3+1$ dimensional inflation,
which is described by the outer spacetime  de Sitter geometry
$ds^2=-dt^2+e^{2Ht} d{\vec x}^2$,
 is driven by an inflaton field $\phi$. Consider vacuum fluctuations of the 
moduli field $\chi_a$ in this geometry. The eigenmode functions
 $\chi_k(t)e^{i\vec k \vec x}$ satisfy 
\begin{equation}\label{mode}
 \ddot \chi_k +3H\dot \chi_k+ \left( \frac{k^2}{a^2}+m^2 \right)\chi_k=0 \ .
\end{equation}
For positive frequency vacuum initial conditions in the far past
its solutions are given in terms of the  Hankel functions
$\chi_k(\tau)=\frac{\sqrt{\pi}}{2} H \vert \tau \vert^{3/2} {\cal H}^{(2)}_{\lambda}(k\tau)$,
with the index $\lambda=\sqrt{\frac{9}{4}-\frac{m^2}{H^2}}$, 
and conformal time $\tau=\int \frac{d t}{a}$.
For very light modes $m_a \ll H$, the amplitude of fluctuations is initially oscillating
and then is frozen out
at the level 
\begin{equation}\label{level}
\chi_k \simeq \frac{H}{\sqrt{2}k^{3/2}}
\end{equation}
when the physical wavelength of the fluctuations exceeds the Hubble radius
$\frac{a(t_k)}{k} > H^{-1}$. This effect means that
 at the end of inflation  there will be an 
inhomogeneous classical scalar field $\chi(\vec x)$ that is
a realization of a random gaussian field, which is a superposition 
of standing waves 
with random phases $\theta_{\vec k}$
and normally distributed  amplitudes   with the variance given by the square of (\ref{level}).
 In the simplest case the spectrum of fluctuations is
scale free, $<\chi^2>=\int d^3k \vert \chi_k\vert^2=\int \frac{d^3k}{k^3}$.
The upper limit of integration corresponds to the latest mode at the end of inflation 
$k_{max} \simeq H$. The lower limit defines the overall variance of fluctuations.
If it is too large, the moduli field may become dominant and play the
role of an
inflaton field \cite{FKL}. We assume that there is no significant value
of $<\chi^2>$, that is to say that there is
 no significant  deformation of the geometry of the compact dimensions
(for instance, bulk gravitons or the radion in braneworlds).
For the sake of generality we may have also an initial homogeneous component
$\chi(t)=<\chi(t, \vec x)>$
whose amplitude is frozen during inflation. 
Notice, however, that varying $m_a$ during inflation
 we may design the spectrum of fluctuations $\chi_k$, independently on the inflaton potential.

For massive modes with $m \leq H$
the effect is absent.
Inflaton field fluctuations $\delta \phi_k(t)$ are generated similarly
provided the effective inflaton mass is smaller than $H$.
It is not mandatory for moduli fields to be light ($m_a \ll H$) during
inflation. Indeed, if one assumes $N=1$ supergravity with the minimal Kahler
potential $K=\frac{\chi^a\chi_a^{*}}{M_P^2}$, then inflation
induces moduli masses $m_a \sim H$ \cite{mass}.  However, 
the same mechanism would make the inflaton field
massive,  which we would like to exclude in order  to save
 standard inflaton fluctuations. This can be done by tuning the Kahler potential
(or using D-term inflation). To continue, we assume that the same cure is
extended to fluctuations of some moduli fields.

To follow the time evolution of the filed $\chi(t, \vec  x)$ after inflation, we assume 
 $\chi(t, \vec  x)=\chi_0(t)+\delta \chi(t, \vec x)$.
There are different possibilities, depending on the stabilization options.
For quadratic moduli potentials  $V(\chi)=\frac{1}{2}m^2 \chi^2$ the
 homogeneous and inhomogeneous parts  are frozen until the decreasing   Hubble parameter
 $H \sim 1/t$   drops below  $m_a$. After that $\chi(t)$ oscillates around its minimum
with decreasing amplitude $\chi(t) \simeq \frac{\chi_0 \cos mt }{a^{3/2}}$, where $\chi_0$
is the amplitude of the initial (homogeneous) displacement.
For runaway moduli potentials the field $\chi$ stays frozen. 

Now we 
can proceed to the next step regarding coupling constant fluctuations
$\frac{\delta \alpha}{\alpha}$,  which we will discuss very soon below.

\section{Spatial Variations of coupling constants}

Next we will consider how spatial variations of moduli fields (which
are present in the time interval
between the end of inflation and 
the time $t_p$ when moduli are pinned down) induce spatial variations of
coupling constants. Effective low energy  string theory,  braneworld models 
and in general $3+1$ effective descriptions  of theories  with extra dimensions 
have coupling constants $\alpha$
that depend on various dilaton/moduli/radion fields, see (\ref{action}) for illustration.
Assuming spatial/temporal variations of these fields,
the coupling constants $\alpha_i$ also vary
\begin{equation}\label{vary}
\frac{\delta \alpha_i}{\alpha_i} \simeq 
\sum_a \frac{\partial \log \alpha_i}{\partial \chi_a} \delta \chi_a (t, \vec x) \ .
\end{equation}
In the standard  model all three couplings are related through the
RG flow equation
\begin{equation}\label{flow}
\frac{1}{\alpha_i}=\frac{1}{\alpha}+\frac{b_i}{2\pi} \log \frac{ E}{\Lambda} \ ,
\end{equation}
where $b_i=(33,1,3)$.
Suppose that $\alpha(\chi_a)$ depends on some of the moduli fields. The
variations of all coupling constants are related $\frac{\delta \alpha_1}{\alpha_1}=
\frac{\delta \alpha_2}{\alpha_2}= \frac{\delta \alpha_3}{\alpha_3}$. 
Notice, however, that it is possible to have variations of the $\alpha$-s decoupled
from each other \cite{Langacker:2001td}.
Temporal variations of $\alpha$  as well as joint time variations of all three
$\alpha_i$-s  in late-time cosmology were considered in  papers
\cite{temp}. Here we introduce a combination of 
 two new elements. First, we concentrate not
on time evolution but on the cosmologically large scale spatial variations of $\alpha$-s.
Second, we consider observational consequences of spatial
$\alpha$ variations not at present, but in the very early universe,
before the epoch when the $\alpha$-s may have been pinned down to their constant
low-energy values.

Spatial variations of the coupling constants in the very early universe lead to the 
generation of large scale isocurvature cosmological perturbations
in the  matter and radiation components consituting  the universe.
Consider   species of particles that are initially in thermal equilibrium
with the hot plasma. Suppose these particles  drop out of equilibrium 
in an expanding universe and their abundance is frozen out
when the rate of their interaction with the hot plasma $\Gamma$ begins to exceed 
the rate of expansion $\Gamma > H$. 
One  example would be weakly interacting neutralinos as CDM particles, which are frozen out  
at $t_{CDM} \sim 10^{-6}$ sec. Another example would be the generation of the baryon
number due to interactions with baryon number violation. 
 Spatial variation of the couplings of these interactions
 $\alpha_i$ at scales larger than the Hubble radius 
results in a slightly different freeze out 
time of these components in different causally connected
 regions (the Hubble patches). This leads to slightly different energy densities
$\rho$ of these components in different Hubble patches
\begin{equation}\label{iso}
\frac{\delta \rho}{\rho} \simeq \frac{\delta \alpha}{\alpha}\sim 
\frac{\partial \log \alpha}{\partial \chi} \delta \chi \ .
\end{equation}
Thus we may have isocurvature fluctuations in the CDM component,
isobaryonic fluctuations etc, modulated by the moduli field(s) fluctuations.
The amplitude of modulated isocurvature fluctuations depends on a few factors,
and our estimate of it bifurcates depending on the way the moduli are pinned down.
Immediately after inflation we estimate 
$\frac{\partial \log \alpha}{\partial \chi} \delta \chi \sim \frac{H}{M_P}$.
If the moduli have a runaway potential, the spatial  variation of $\alpha$ stays
at its initial value. In this case all species of relic particles which are frozen 
out of thermal equilibrium will have isocurvature fluctuations
$\frac{\delta \rho}{\rho} $ defined by the initial value of the $\alpha$ variation.

 For moduli potentials with a minimum, both the homogeneous
component $\chi$ and its fluctuations $\chi_k$ at some point begin to oscillate.
In the limiting case of a vanishing homogeneous moduli component $\chi(t)=0$, the amplitude of
the inhomogeneous fluctuations $\delta \chi$ decreases with time $\delta \chi \sim 1/a^{3/2}$,
as does the $\delta \alpha$ variation. In this case, again, the $\frac{\delta \rho}{\rho} $
 amplitude depends on how late the relic component is freezed out
and  effect can  be negligible.
Interesting modulated isocurvature perturbations may arise in baryonic component
if baryon assymetry is generated very soon after inflation.
Interesting modulated isocurvature perturbations in CDM component are possible
if CDM particle frezed out very soon after inflation.
If moduli responsible for both curvature and isocurvature fluctuations,
they can be correlated. Correlated curvature and isocurvature fluctuations
can produce interesting combine effects \cite{Gordon:2000hv},
although there are observational constraint on isocurvature fluctuations
\cite{Bennett:2003bz}

\section{Modulated curvature perturbations}

This mechanism of generation of modulated perturbations works wherever
different species are
out of equilibrium and  couplings are important,
and coupling constants are inhomogeneous at the scales larger than
the horizon.
 Let us apply it to the
origin of matter from (p)reheating after inflation.
Indeed, in the theory of inflation the subsequent  radiation domination stage 
takes place after the decay of the inflaton field $\phi$ due to its
coupling with other particles (for instance, in the process of preheating after inflation
\cite{KLS94}). Suppose the coupling is modulated by the
spatial inhomogeneities of the moduli field. Then the transition from
the inflaton dominated
to the radiation dominated stage occurs at slightly different times in different Hubble
patches, and we will have modulated isocurvature fluctuations in radiation after
inflation. Since radiation becomes gravitationally dominant, its
isocurvature fluctuations will turn into curvature fluctuations soon after inflation.

There is an interesting limiting case of this scenario. Let us 
assume that   the scalar fluctuations
from inflation are suppressed 
(say, by our choice of the parameters of the inflaton potential).
What will be left will be the moduli fluctuations,  which will trigger modulated
 isocurvature fluctuations in radiation.
These will be transferred into curvature fluctuations of the amplitude
$\sim \frac{\delta \chi_a}{\chi_a}$.
 This is an alternative mechanism 
to obtain scale free curvature fluctuations from inflation, which does not invoke
the details of the inflaton potential. We will call these fluctuations 
modulated curvature fluctuations. 
This idea is easily illistrated in  perturbative reheating after inflation.
Indeed, let $\Gamma_{tot}$ be the rate of inflaton decay into radiation.
The reheating temperature is $T_r \simeq \sqrt{\Gamma_{tot} M_p}$.
Variations in  $\Gamma_{tot}$ due to the variation of the coupling constan
results in  entropy fluctuations $\delta T_r$, which convert
into curvature fluctuations after inflaton decay.

Typically, however, reheating goes through a  non-perturbative, non-linear
preheating stage \cite{Kofman:1997yn} and the
dependence of the reheat temperature on
the couplings is rather complicated  \cite{Felder:2000hr}.
However, on  general grounds we still  expect that at large scales
spatial variations in 
the final reheat temperature  $\frac{\delta T_r}{T_r}$ will be linear
with respect to the 
 couplings, $\frac{\delta T_r}{T_r} \simeq \frac{\delta \alpha}{\alpha}$.

\section{Discussion}

We suggest a new mechanism for generating cosmological fluctuations
from inflation. The main idea is to modulate coupling constants 
by large scale spatial variations of the moduli fields.
We can obtain modulated isocurvature fluctuations and modulated curvature
fluctuations and even a correlated mixture of them.
This may require significant tuning.
However, there is a particularly interesting case of generation of
modulated curvature fluctuations, which does not require tuning.

This mechanism has some common features with 
 the curvaton scenario \cite{Lyth:2001nq}.
However, in our scheme there  are no
 additional  assumptions about the decay of the curvaton,
since the inflaton field decays anyway.
In this respect, the model of modulated curvature
fluctuations is more economic. 
If the inflaton decay goes gravitationally, then our mechanism is related to
spatial variation of the effective Planckian mass (dilaton),
considered in \cite{Garcia-Bellido:1993wn}.
The mechanism of curvature fluctuations due to
 coupling constant variations similar to ours was independently
suggested in \cite{DGZ}.

\section*{Acknowledgments}
I am grateful to    G.~Felder,             
          N.~Kaloper and.~Linde 
 for valuable discussions. This
research was supported NSERC and CIAR.



\begin{references}

\bibitem{scalar} V.~F.~Mukhanov and G.~V.~Chibisov,
``Quantum Fluctuation And 'Nonsingular' Universe,''
JETP Lett.\  {\bf 33}, 532 (1981)
[Pisma Zh.\ Eksp.\ Teor.\ Fiz.\  {\bf 33}, 549 (1981)]; S.~W.~Hawking,
``The Development Of Irregularities In A Single Bubble Inflationary Universe,''
Phys.\ Lett.\ B {\bf 115}, 295 (1982); A.~A.~Starobinsky,
``Dynamics Of Phase Transition In The New Inflationary Universe Scenario And Gen
eration Of Perturbations,''
Phys.\ Lett.\ B {\bf 117}, 175 (1982); A.~H.~Guth and S.~Y.~Pi,
``Fluctuations In The New Inflationary Universe,''
Phys.\ Rev.\ Lett.\  {\bf 49}, 1110 (1982); J.~M.~Bardeen, P.~J.~Steinhardt and
M.~S.~Turner,
``Spontaneous Creation Of Almost Scale - Free Density Perturbations In An Inflat
ionary Universe,''
Phys.\ Rev.\ D {\bf 28}, 679 (1983).

\bibitem{tensor}  A. Starobinsky, ``Relic gravitational radiation spectrum
 and initial state of the universe'', JETP Lett {\bf 30} (1979) 682.

\bibitem{cosmo} first reported in my  plenary talk at Cosmo03, Chicago, September 2002;
available at http://pancake.uchicago.edu/~cosmo02/

\bibitem{dine} M.~Dine, ``Towards solution of the moduli problems of string cosmology'',
[arXiv:hep-ph/0002047].

\bibitem{brane} 
C.~Csaki, M.~L.~Graesser and G.~D.~Kribs,
``Radion dynamics and electroweak physics,''
Phys.\ Rev.\ D {\bf 63}, 065002 (2001)
[arXiv:hep-th/0008151].

\bibitem{strings} J.~Polchinski, {\it String Theory}, vol.1,  Cambridge University Press 1998.


\bibitem{FKL}
G.~N.~Felder, L.~Kofman and A.~D.~Linde,
``Gravitational particle production and the moduli problem,''
JHEP {\bf 0002}, 027 (2000)
[arXiv:hep-ph/9909508].

\bibitem{mass}
M.~Dine, L.~Randall and S.~Thomas,
``Supersymmetry breaking in the early universe,''
Phys.\ Rev.\ Lett.\  {\bf 75}, 398 (1995)
[arXiv:hep-ph/9503303].

\bibitem{Langacker:2001td}
P.~Langacker, G.~Segre and M.~J.~Strassler,
``Implications of gauge unification for time variation of
 the fine  structure constant,''
Phys.\ Lett.\ B {\bf 528}, 121 (2002)
[arXiv:hep-ph/0112233].

\bibitem{Gordon:2000hv}
C.~Gordon, D.~Wands, B.~A.~Bassett and R.~Maartens,
``Adiabatic and entropy perturbations from inflation,''
Phys.\ Rev.\ D {\bf 63}, 023506 (2001)
[arXiv:astro-ph/0009131].

\bibitem{Bennett:2003bz}
C.~L.~Bennett {\it et al.},
``First Year Wilkinson Microwave Anisotropy Probe (WMAP)
 Observations: Preliminary Maps and Basic Results,''
arXiv:astro-ph/0302207.


\bibitem{KLS94}
L.~Kofman, A.~D.~Linde and A.~A.~Starobinsky,
Phys.\ Rev.\ Lett.\  {\bf 73}, 3195 (1994)
[arXiv:hep-th/9405187].

\bibitem{Kofman:1997yn}
L.~Kofman, A.~D.~Linde and A.~A.~Starobinsky,
``Towards the theory of reheating after inflation,''
Phys.\ Rev.\ D {\bf 56}, 3258 (1997)
[arXiv:hep-ph/9704452].

\bibitem{Felder:2000hr}
G.~N.~Felder and L.~Kofman,
``The development of equilibrium after preheating,''
Phys.\ Rev.\ D {\bf 63}, 103503 (2001)
[arXiv:hep-ph/0011160];
R.~Micha and I.~I.~Tkachev,
``Relativistic turbulence: A long way from preheating to equilibrium,''
arXiv:hep-ph/0210202.

\bibitem{Lyth:2001nq}
D.~H.~Lyth and D.~Wands,
``Generating the curvature perturbation without an inflaton,''
Phys.\ Lett.\ B {\bf 524}, 5 (2002)
[arXiv:hep-ph/0110002].

\bibitem{Garcia-Bellido:1993wn}
J.~Garcia-Bellido, A.~D.~Linde and D.~A.~Linde,
``Fluctuations of the gravitational constant in the
 inflationary Brans-Dicke cosmology,''
Phys.\ Rev.\ D {\bf 50}, 730 (1994)
[arXiv:astro-ph/9312039].

\bibitem{DGZ} G.~Dvali, A.~Gruzinov, M.~Zaldarriaga,
``A new mechanism for generating density perturbations
from inflation'', [arXiv:astro-ph/0303591].


\end{references}
\end{document}